\documentclass[12pt,preprint]{aastex}
\shorttitle{}
\shortauthors{Nesvorn\'y et al.}
\begin{document}
\title{CAPTURE OF IRREGULAR SATELLITES AT JUPITER}
\author{David Nesvorn\'y$^1$, David Vokrouhlick\'y$^{1,2}$, Rogerio Deienno$^{1,3}$}
\affil{(1) Department of Space Studies, Southwest Research Institute, 1050 Walnut St., \\Suite 300, 
Boulder, CO 80302, USA} 
\affil{(2) Institute of Astronomy, Charles University, V Hole\v{s}ovi\v{c}k\'ach 2, \\
180 00 Prague 8, Czech Republic}
\affil{(3) Division of Space Mechanics and Control, National Institute of Space Research,\\
S\~ao Jos\'e dos Campos, SP 12227-010, Brazil }


\begin{abstract}
The irregular satellites of outer planets are thought to have been captured from heliocentric orbits.
The exact nature of the capture process, however, remains uncertain. We examine the possibility 
that irregular satellites were captured from the planetesimal disk during the early Solar System 
instability when encounters between the outer planets occurred (Nesvorn\'y, Vokrouhlick\'y
\& Morbidelli 2007, AJ 133; hereafter NVM07). NVM07 already showed that the irregular satellites
of Saturn, Uranus and Neptune were plausibly captured during planetary encounters. Here we find that 
the current instability models present favorable conditions for capture of irregular satellites at 
Jupiter as well, mainly because Jupiter undergoes a phase of close encounters with an ice giant. We 
show that the orbital distribution of bodies captured during planetary encounters provides a good 
match to the observed distribution of irregular satellites at Jupiter. The capture efficiency for 
each particle in the original transplanetary disk is found to be (1.3-$3.6)\times10^{-8}$. This is 
roughly enough to explain the observed population of jovian irregular moons. We also confirm NVM07's 
results for the irregular satellites of Saturn, Uranus and Neptune.
\end{abstract}

\section{Introduction}
The four outer planets in the Solar System have a significant population of satellites with
large, elongated, inclined, and often retrograde orbits (see, e.g., Nicholson et al. 2008 for a 
review). These {\it irregular} satellites are thought to have been captured by planets from 
heliocentric orbits, because the circumplanetary disk processes that are thought to be 
responsible for the formation of the regular satellites are not expected to lead to the orbital 
characteristics of the irregular satellites. Many capture mechanisms have been proposed in the 
past (e.g., Colombo \& Franklin 1971, Heppenheimer \& Porco 1977, Pollack et al. 1979). Here we 
consider the capture mechanism suggested in NVM07. 

NVM07 proposed that the irregular satellites were captured around giant planets when the orbital 
instability in the outer Solar System triggered a phase of close encounters of Uranus and Neptune 
to Jupiter and Saturn. Each encounter involves a complex gravitational interaction of a gas giant,
an ice giant, and numerous background planetesimals. If the encounter geometry is right, the 
trajectory of a background planetesimal can be influenced in such a way that the planetesimal 
ends up on a bound orbit around one of the planets, where it remains permanently trapped when 
planets move away from each other. 
  
To illustrate this concept, NVM07 used the model of planetary instability and migration proposed by 
Tsiganis et al. (2005; hereafter the original Nice model or ONM for short). In the ONM, the outer 
planets were placed between 5 and 18 AU. The instability was triggered when Jupiter and Saturn
migrated over their mutual 2:1 resonance. Following this event, the orbits of Uranus and Neptune became 
Saturn-crossing, Uranus and Neptune were scattered out by Saturn, and these planets stabilized 
and migrated to their current locations by gravitationally interacting with an outer planetesimal disk. 

The encounters between planets in the ONM remove distant satellites that may have initially 
formed at Saturn, Uranus and Neptune by gas-assisted capture (or via a different mechanism). A new 
generation of satellites, however, can be captured from the background planetesimal disk during 
planetary encounters. By modeling this process, NVM07 showed that the capture efficiency and orbital 
distribution of bodies captured at Saturn, Uranus and Neptune are consistent with observations.

Because Jupiter does not generally participate in planetary encounters in the ONM, however, the proposed 
mechanism was not expected to produce the irregular satellites of Jupiter. Yet, the population of 
irregular satellites at Jupiter does not seem to be notably different from those at Saturn, Uranus or 
Neptune (Jewitt \& Haghighipour 2007). This problem may be rooted in our incomplete understanding of 
the precise ways the outer planets formed and evolved by interacting with the planetesimal disk.

The currently favored model of the instability in the outer Solar System is that of {\it jumping 
Jupiter}. This model was proposed to avoid problems with excessive excitation of the terrestrial 
planets that would occur if Jupiter and Saturn slowly migrated past the 2:1 MMR (Brasser et al. 2009, 
2013; Agnor \& Lin 2012), to explain the dynamical structure of the asteroid belt (Minton \& Malhotra 
2009, Morbidelli et al. 2010), and to obtain a correct distribution of secular modes for Jupiter
and Saturn (Morbidelli et al. 2009a). 

In the jumping-Jupiter model, Jupiter undergoes a series of encounters with Uranus, Neptune or the 
ejected ice giant (Nesvorn\'y 2011, Batygin et al. 2012, Nesvorn\'y \& Morbidelli 2012; hereafter NM12). 
This may resolve the problem with the NVM12 results discussed above, because the irregular satellites 
of Jupiter could have been captured during these encounters. Here we study the capture of irregular 
satellites at Jupiter in the instability models taken from NM12. Our methods are explained in Section~2. 
The capture efficiency and orbital distribution of captured bodies are reported in Section~3. Section 
4 concludes this paper.
\section{Capture Simulations}
\begin{figure}
\epsscale{0.5}
\plotone{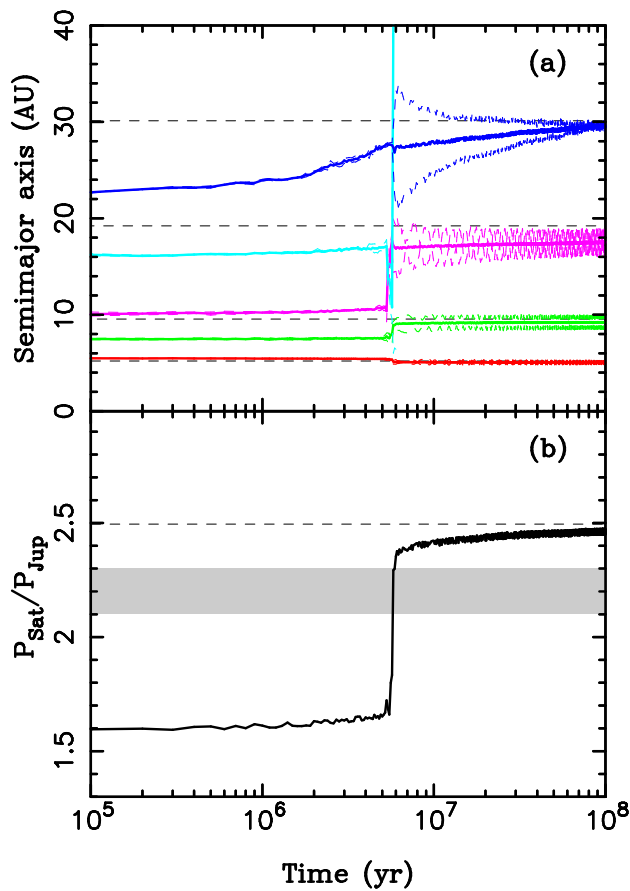}
\caption{Orbital histories of the outer planets in Case 1. The planets were started in the (3:2,3:2,2:1,3:2) resonant chain, 
and $M_{\rm disk}=20$ M$_{\rm Earth}$. (a) The semimajor axes (solid lines), and perihelion and aphelion distances (dashed 
lines) of each planet's orbit.  The black dashed lines show the semimajor axes of planets in the present Solar System. (b) 
The period ratio $P_{\rm Sat}/P_{\rm Jup}$. The dashed line shows $P_{\rm Sat}/P_{\rm Jup}=2.49$, corresponding to the period 
ratio in the present Solar System. The shaded area approximately denotes the zone where the secular resonances with the 
terrestrial planets occur.}
\label{case1}
\end{figure}

Here we work with three cases taken from NM12. Their properties were illustrated in Nesvorn\'y et al. 
(2013, their Figures 1-4). They were selected from NM12 as three representative instability simulations 
that satisfy the success criteria defined in NM12 (i.e., generate correct orbits of the outer planets, 
avoid excessive excitation of the inner planets, and produce correct amplitude of the proper 
eccentricity mode of Jupiter's orbit). In all three cases, the Solar System was assumed to 
have five giant planets initially 
(Jupiter, Saturn and three ice giants). This is because NM12 showed that having five planets initially 
is convenient to obtain jumping Jupiter and satisfy constraints. The third ice giant with the mass comparable 
to that of Uranus or Neptune is ejected into interstellar space during the instability (Figure \ref{case1};
see also Nesvorn\'y 2011, Batygin et al. 2012). A shared property of the selected runs is that Jupiter 
undergoes a series of encounters with the ejected ice giant.  

NM12's simulations were performed using the symplectic integrator known as {\tt SyMBA} (Duncan et al. 1998). 
The outer planetesimal disk in NM12 was represented by up to 10,000 particles distributed in an annulus with 
the inner edge at $r_{\rm in}$ and the outer edge at $r_{\rm out}$, where $r_{\rm in}=20$-24 AU, with the exact 
value depending on the specific case, and $r_{\rm out}=30$ AU. The surface density of particles in the annulus 
was set to $\Sigma \propto 1/r$, where $r$ is the radial distance from the Sun. The eccentricities and 
inclinations of particles were set to be zero. 

Because the capture probabilities are expected to be very low ($\sim 10^{-8}$ for each disk particle; NVM07), 
the original number of disk particles in NM12 was largely insufficient to detect satellite capture directly. 
Instead, to be able to deal with this very low probability, we adopt the the methodology developed in NVM07.
This involves a three step process, where different numerical integrators are used to model the initial stage 
of the outer disk dispersal, planetesimal dynamics during the instability, and the capture process itself.
At each transition, the results of the previous step are used to set up the initial conditions for the 
next step. The main goal of this procedure is to increase the statistics.

We start by characterizing the overall spatial density and orbital distribution of planetesimals as a function 
of time and location in the disk. This is done as follows. The selected NM12 runs are repeated with 
{\tt SyMBA}, using the same initial conditions for planets and planetesimals, and recording the planetary 
orbits at 1-yr time intervals. We then perform a second set of integrations with our modified version of the  
{\tt swift\_rmvs3} integrator (Levison \& Duncan 1994), where the planetary orbits are read from the file 
recorded by {\tt SyMBA} and are interpolated to the required sub sampling (generally 0.25 yr, which is the 
integration time step used here). The interpolation is done in Cartesian coordinates 
(Nesvorn\'y et al. 2013).\footnote{First, the planets are forward propagated 
on the ideal Keplerian orbits starting from the positions and velocities recorded by SyMBA at the beginning 
of each 1-yr interval. Second, the SyMBA position and velocities at the end of each 1-yr interval are
propagated backward (again on the ideal Keplerian orbits). We then calculate a weighted
mean of these two Keplerian trajectories for each planet so that progressively more (less)
weight is given to the backward (forward) trajectory as time approaches the end of the 1-yr
interval. We verified that this interpolation method produces insignificant errors.} This assures that 
the orbital evolution of planets in these new integrations is practically the same (up to small errors 
caused by the interpolation routine) as in the original {\tt SyMBA} runs.

\begin{figure}
\epsscale{0.7}
\plotone{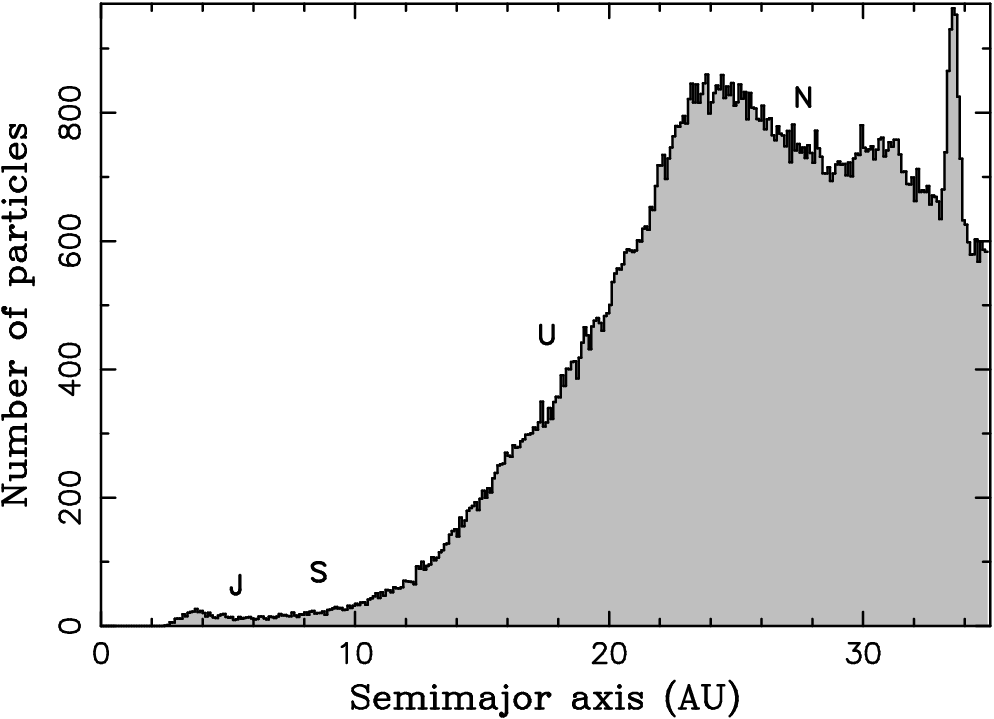}
\caption{The semimajor axis distribution of disk particles in Case 1 at the time of the first encounter between
Jupiter and an ice giant. The labels denote the semimajor axes of planets at this instant. The peak in the 
distribution at $\simeq$33.5 AU corresponds to particles captured in the 3:2 MMR with Neptune.}
\label{disk}
\end{figure}

The {\tt swift\_rmvs3} jobs were launched on different CPUs, with each CPU computing the orbital evolution 
of a large number of disk particles. The particles were considered massless so that they do not interfere 
with the interpolation routine (this is why {\tt swift\_rmvs3}, and not {\tt SyMBA}, was used at this 
stage). The initial orbital distribution of each particle set was chosen to respect 
the initial distribution of disk particles in the original NM12 simulation, but differed in details (e.g., the 
initial mean longitudes of particles were random), so that each set behaved like an independent statistical sample. 
This allowed us to build up good statistics. In total, we followed $N_{\rm disk} = 300,000$ disk particles in 
each of the three selected cases. Figure \ref{disk} illustrates the semimajor axis distribution of disk 
particles in Case 1.

Our {\tt SyMBA} integrations were also used to record all encounters between planets with encounter 
distance $r<R_{\rm H,1}+R_{\rm H,2}$, where $R_{\rm H,1}$ and $R_{\rm H,2}$ are the Hill radii of the two planets
having an encounter. Then, in the {\tt swift\_rmvs3} runs with the increased resolution ($N_{\rm disk} = 300,000$), 
we drew a sphere with radius 3 AU around each encounter and characterized the orbital distribution of disk 
particles in this `encounter zone'. This was done individually for each encounter. In the following, the fraction 
of disk particles in the encounter zone will be denoted by $f_{{\rm 3AU},i}=N_{{\rm 3AU},i}/N_{\rm disk}$, 
where $N_{{\rm 3AU},i}$ is the number of disk particles in the zone and index $i$ runs over encounters. 
Typically, we find that $f_{{\rm 3AU},i}\simeq4\times10^{-4}$ for Jupiter.  

In the final step, we used the Bulirsch-Stoer code that we adapted to this problem in NVM07, and followed
the planets and disk particles through a sequence of encounters.  The timing and geometries of planetary 
encounters were taken from the {\tt SyMBA} integrations. For each encounter, the two planets were first 
integrated backward from the moment of their closest encounter until their separation increased to $3 
(R_{\rm H,1}+R_{\rm H,2})$. We then included $N_{\rm test}=4\times10^7$ test particles with the orbital distribution 
respecting that of the disk particles in the encounter zone. The mean longitudes of the particles were set 
such that, if particles were propagated on the strictly Keplerian heliocentric orbits, they would end up in 
the 3 AU encounter zone at the time of the closest approach between planets. The forward integrations with 
planets and test particles were run up to $t=2 T$, where $t=T$ corresponds to the time of the closest 
approach. 

After each planetary encounter, the Hill spheres of planets were searched for captured satellites. The orbits 
of the identified satellites were integrated to the next encounter (following the time sequence of encounters
recorded by {\tt SyMBA}. The unstable orbits were removed. The 
stable satellites were kept and included in the next encounter. By iterating the procedure over all encounters, 
we obtained the state of the satellite swarm at each planet after the last planetary encounter. We then followed 
the orbits of these satellites with a symplectic integrator for additional 100 Myr to account for long term 
instabilities and removal by collisions with large regular moons.\footnote{The large regular satellites are 
assumed to have formed at this stage. This should be obvious, because the regular satellites presumably formed 
in the circumplanetary disk within the first $\sim10$ Myr after the condensation of first solar system solids, 
while here we are describing events that occurred much later.} It is not necessary to follow orbits over 
gigayear timescales, because the number of removed satellites at late times is minimal. The final population 
of stable satellites is discussed below. 

For consistency, we also evaluated the dynamical survival of {\it regular} moons during the planetary encounters 
in three cases considered here. The results of this part of the work will be reported in Deienno et al. 
(2014, in preparation).
\section{Results}
Here we describe the results of the numerical integrations described above. The mechanism of capture is 
discussed in \S3.1. We then examine the orbital distribution of captured irregular satellites and compare 
it with observations (\S3.2). The capture probability and implications for the size distribution of planetesimals 
in the outer disk are discussed in \S3.3. Here we focus on the irregular satellites of Jupiter, because they are the
most troublesome case given that their capture does not occur in the ONM (NVM07). The irregular satellites 
of other planets are briefly discussed in \S3.4. 
\subsection{Captured Satellites}
\begin{figure}
\epsscale{0.323}
\plotone{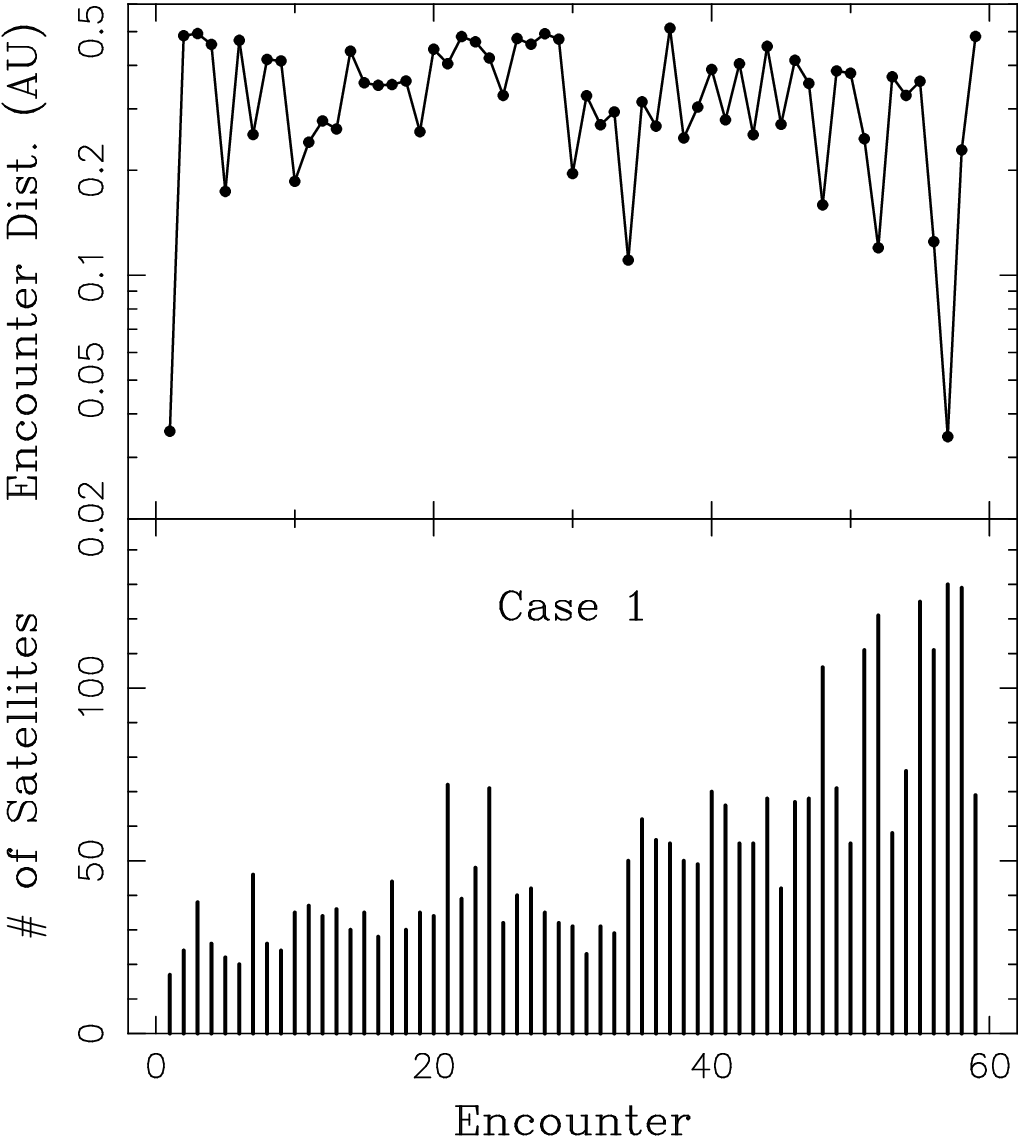}
\epsscale{0.30}
\plotone{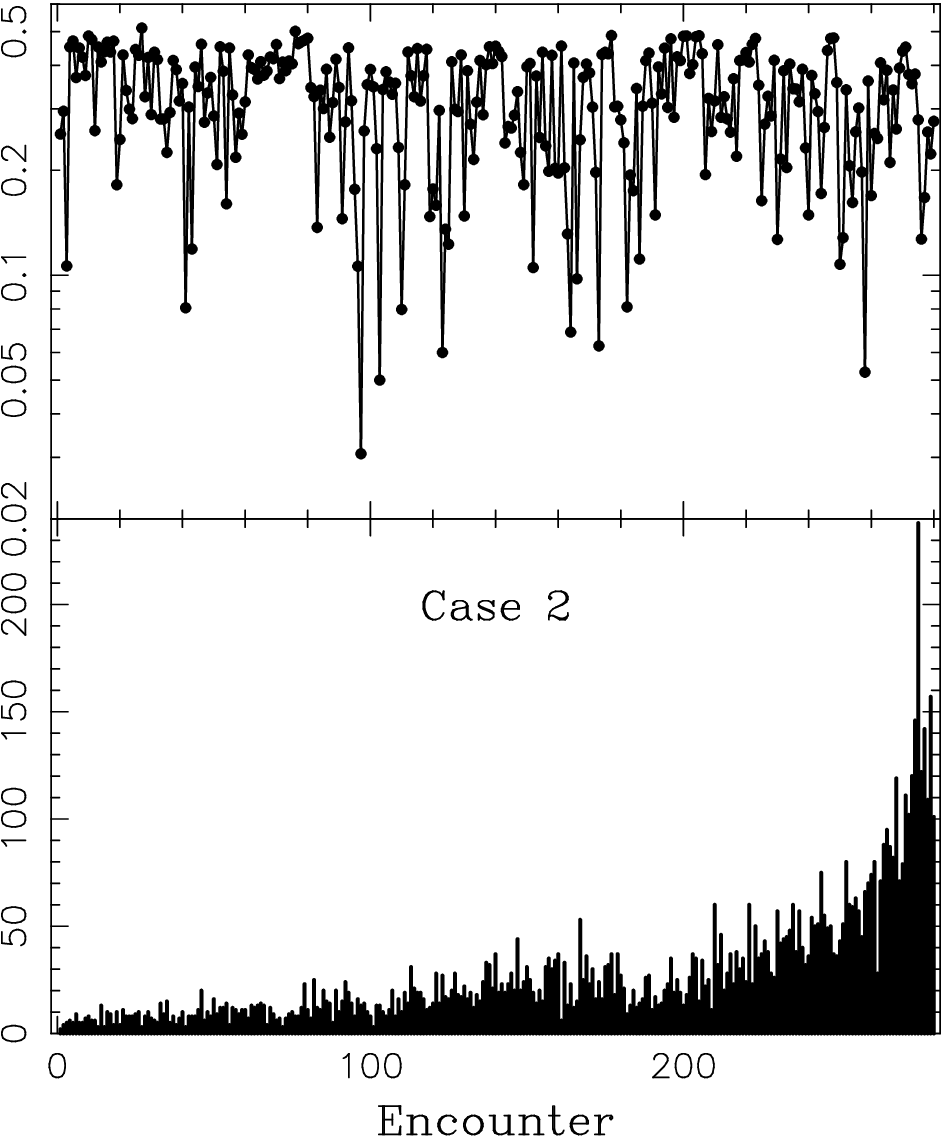}
\plotone{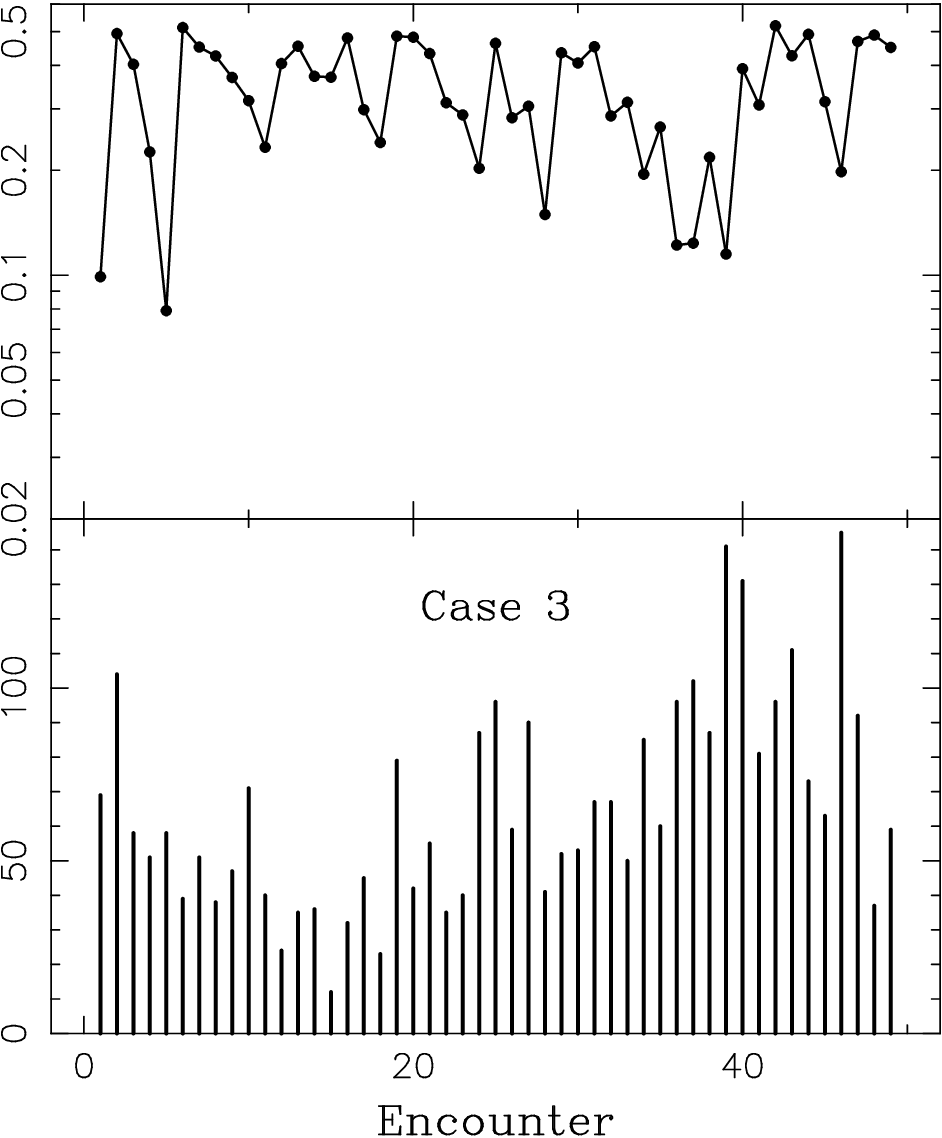}
\caption{The history of encounters (top) and number of satellites (bottom) captured at Jupiter in
Cases 1, 2 and 3. Only the stable satellites, i.e., those that survive all subsequent encounters and 
100 Myr past the encounter stage, are shown in the bottom histograms.}
\label{encs}
\end{figure}

While the global evolution of planets was similar in the three selected cases (Nesvorn\'y et al. 2013; their 
Figures 1-3), the history of Jupiter's encounters with an ice giant varied from case to case. These differences 
can be important for satellite capture and is why different cases were considered in the first place
(Figure \ref{encs}). In Case 1, there were 59 encounters of Jupiter with an ice giant occurring over an 
interval of 200 kyr. Two of these encounters, one near the beginning and one near the end of the scattering phase, 
were very deep. In Case 2, the scattering phase of Jupiter was richer, with 280 recorded encounters, and lasted 
over 300 kyr. In contrast, Case 3 showed a relatively poor history of Jupiter's encounters lasting 40 kyr only.

Figure \ref{encs} shows the number of satellites captured during different encounters at Jupiter. This plot 
illustrates, as already reported in NVM07, that satellites can be captured at nearly all close encounters. 
The satellites captured during early encounters, however, are often eliminated by the subsequent encounters.
The satellites that survive subsequent encounters are typically the ones that were captured on closely bound
orbits (semimajor axis $a\lesssim 0.1$ AU). The population of satellites builds up over time as more and more 
encounters contribute. The largest number of satellites was produced in Case 2, which is expected, 
because Case 2 has the largest number of encounters. 

The satellites are captured in our simulations when the presence of the ice giant influences the initially
hyperbolic fly-by of a test particle near Jupiter. Capture happens even during relatively distant encounters 
when the Hill spheres of the two planets barely overlap (Figure \ref{encs}). In this case, particles passing 
near the ice giant cannot be captured on bound orbits about Jupiter, because they will enter Jupiter's Hill 
sphere on a hyperbolic trajectory. Instead, the gravity of the ice giant influences the trajectories 
of bodies passing deeply inside the Hill sphere of Jupiter, and changes them, if the encounter geometry
is right, such that they become bound. 

\subsection{Orbital Distribution}
\begin{figure}
\epsscale{0.6}
\plotone{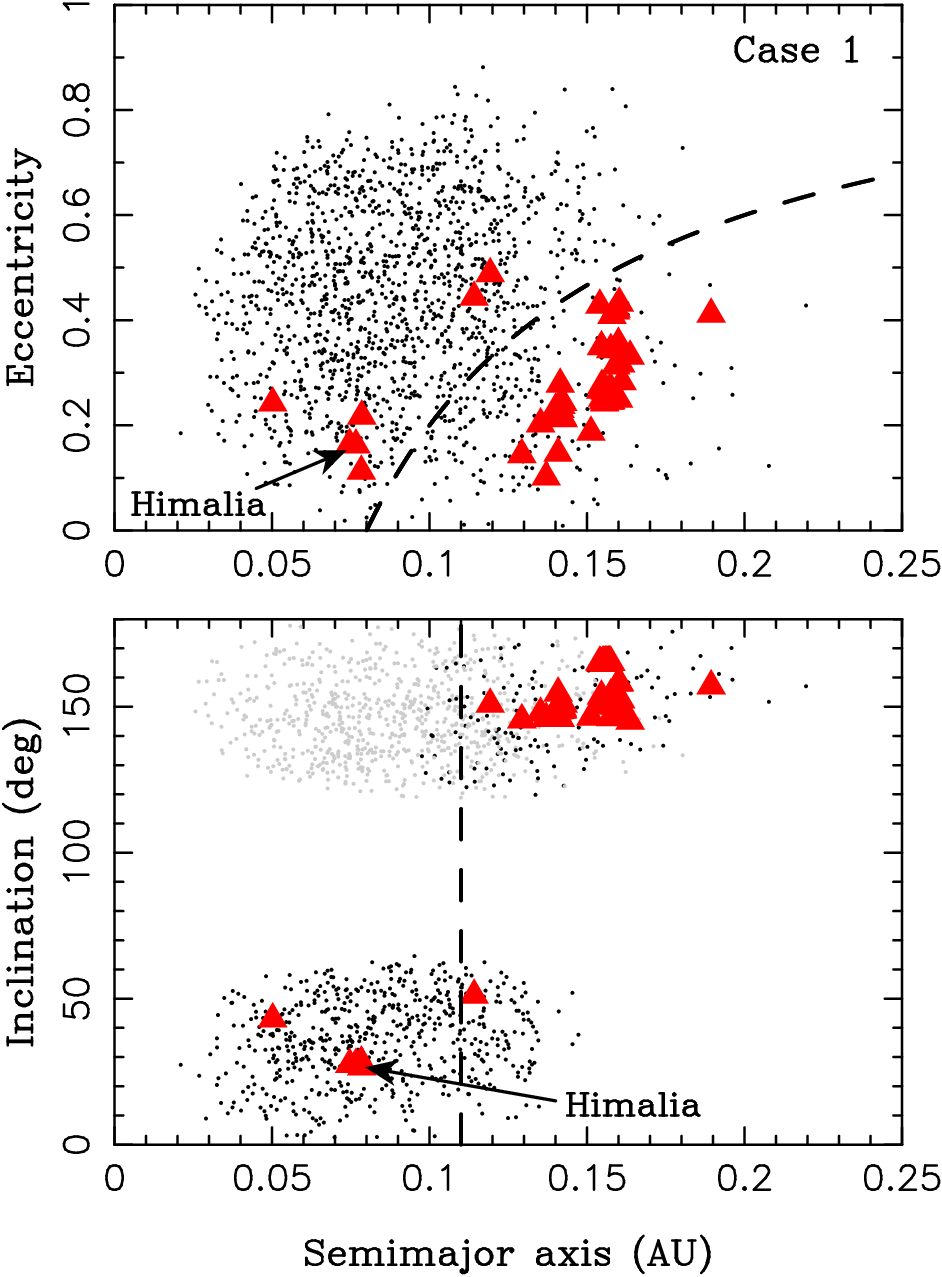}
\caption{The orbits of satellites captured in Case 1 (dots) and known irregular satellites at Jupiter 
(red triangles). The dashed line in the top panel denotes $q=a(1-e)=0.08$~AU, which is an approximate
limit below which the population of small retrograde satellites becomes strongly depleted by collisions 
with Himalia (Nesvorn\'y et al. 2003). The depleted orbits with $i>90^\circ$ and $q<0.08$~AU are shown by 
grey dots in the bottom panel. The dashed line in the bottom panel shows the boundary value $a=0.11$ AU 
below which the collisions with Himalia whould remove more than 50\% of small retrograde satellites
(Nesvorn\'y et al. 2003).}
\label{jc1}
\end{figure}

The orbital distribution of satellites captured at Jupiter in Case 1 is shown in Figure \ref{jc1}. The 
orbital distributions obtained in Cases 2 and 3 are very similar to that of Case 1 and are not shown here. 
The semimajor axis of captured bodies ranges between 0.03 and 0.14 AU for prograde orbits, and 0.03 and 
0.2 AU for retrograde orbits. The different outer extensions are dictated by different stability limits of 
the prograde and retrograde orbits (e.g., Nesvorn\'y et al. 2003). The eccentricity distribution covers the 
whole range of values between 0 and $\simeq$0.7. The orbits with $q=a(1-e)<0.015$ AU were removed, because
of collisions with the Galilean satellites. The orbits with inclinations $60^\circ<i<120^\circ$ are strongly 
affected by the Kozai resonance (e.g., Carruba et al. 2002) and are removed as well once they reach 
$q<0.015$ AU. 

The comparison with orbits of known irregular satellites at Jupiter is satisfactory. The known prograde
satellites fall well within the range of the model distribution. The distribution of known retrograde 
orbits is somewhat discordant with the model distribution in that all known retrograde satellites have
$a>0.11$ AU, while the semimajor axes of captured bodies extend down to 0.03 AU (Figure \ref{jc1},
bottom panel). This can be understood, because the model distribution shown here does not account
for collisions with the large prograde moon Himalia (mean radius $R\simeq85$ km). Once these collisions
are taken into account, it becomes apparent that the population of small retrograde satellites with 
$q<0.08$ AU must have been strongly depleted over gigayear timescales. For example, Nesvorn\'y et al. 
(2003) estimated that {\it most} retrograde satellites with $a<0.11$ AU should be removed by collisions 
with Himalia.

We therefore conclude that the orbital distribution of Jovian irregular satellites was shaped in a rather 
intricate way by the capture process, long-term instabilities, collisions with the regular moons, and 
collisions of the irregular moons themselves (see also Bottke et al. 2010). The final orbital 
distributions are found to be relatively insensitive to the detail history of planetary encounters in 
the NM12 models, because all three cases considered here give similar orbital distributions. This result
provides support to the NM12 models considered here, because these models were not selected with any
a priori knowledge of what to expect.    
\subsection{Capture Probability}
The probability of capture for each disk particle can be written as
\begin{equation}
P_{\rm cap} = \sum_{i=1}^{N_{\rm enc}} f_{{\rm 3AU},i} {N_{{\rm cap},i} \over N_{\rm test}} \ , 
\end{equation}
where index $i$ goes over individual planetary encounters, $N_{\rm enc}$ is the total number of recorded 
encounters, $f_{{\rm 3AU},i}$ is the fraction of disk particles in the 3-AU-radius encounter sphere (see 
Section 2), $N_{{\rm cap},i}$ is the number of {\it stable} satellites captured during encounter $i$, 
and $N_{\rm test}=4\times10^7$ is the number of test particles injected into the encounter sphere. 
The values of these parameters for our three cases are reported in Table 1. We find that $P_{\rm cap} = 
(1.3$-$3.6)\times10^{-8}$. This is 1.5-4.2 times more than $P_{\rm cap}$ found in one isolated case in 
NVM07. 

In Nesvorn\'y et al. (2013), we used the very same three instability cases from NM12 considered here to 
study capture of Jupiter Trojans. The results implied, when calibrated from the observed population of 
Jupiter Trojans, that the outer planetesimal disk contained $\simeq(3$-$4)\times10^7$ bodies with effective 
radius $R>40$ km. Using this calibration and the capture probability computed here, we find that roughly 
one irregular satellite with $R>40$~km is expected to be captured at Jupiter. 
For comparison, two largest irregular moons, Himalia and Elara, have $R\sim85$ km and $R\sim43$ km, 
respectively (the third largest is Pasiphae with estimated $R\sim30$ km). This comparison is encouraging 
because it shows that the calculations are roughly consistent with the observed number of satellites.

\begin{figure}
\epsscale{0.8}
\plotone{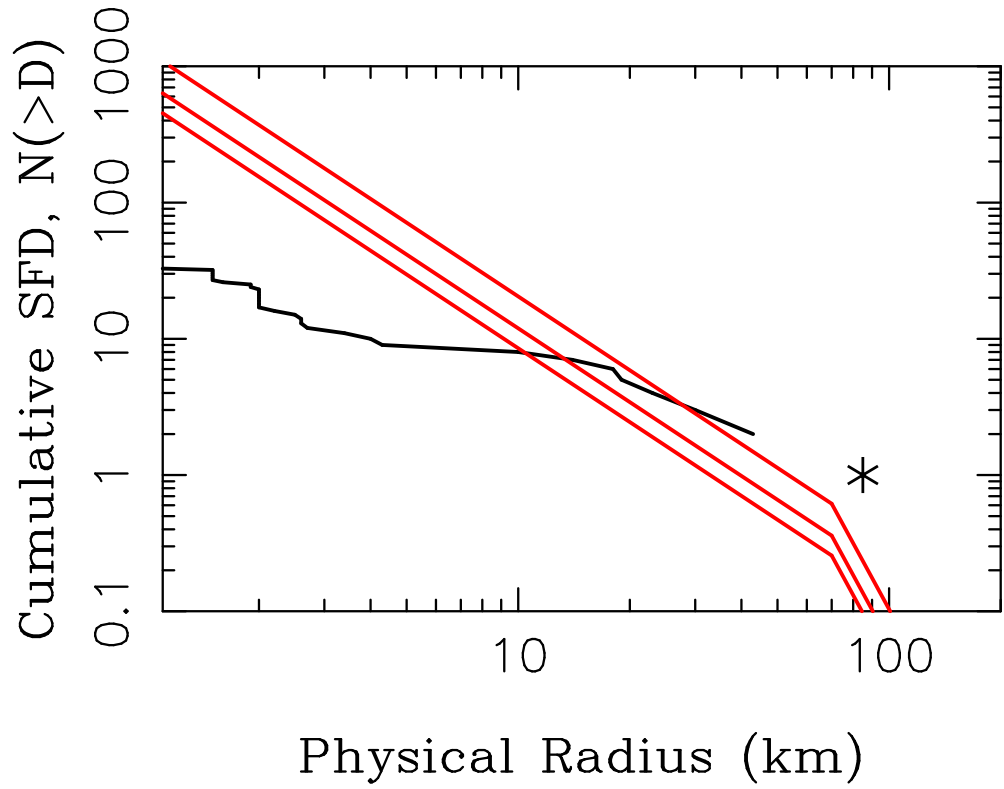}
\caption{The cumulative size distribution of known irregular satellites at Jupiter (black line).
The largest satellite, Himalia, is shown by an asterisc. The three red lines show the distributions
expected from our model in Cases 2, 1 and 3 (from top to bottom).}
\label{sfd}
\end{figure}

Figure \ref{sfd} shows the results in a plot. Here we assume that the shape of the Size Frequency 
Distribution  (SFD) of Jupiter Trojans represents a good proxy for the SFD of planetesimals in the 
outer disk (Morbidelli et al. 2009b). If so, the SFD of disk planetesimals can be approximated by 
$N(>R)=N_0(R/R_*)^{-q_1}$ for $R>R_*=70$ km and $N(>R)=N_0(R/R_*)^{-q_2}$ for $R<R_*$, where 
$q_1\simeq5$ and $q_2\simeq1.8$ (e.g., Fraser et al. 2014). The normalization constant $N_0$ is set 
such that $N(>40\,{\rm km})\simeq(3$-$4)\times10^7$, as found in Nesvorn\'y et al. (2013)
by calibrating the SFD from Jupiter Trojans. We than convolve this SFD with the capture probability 
reported above.
  
The results shown in Figure \ref{sfd} indicate that the capture process studied here is capable of capturing 
a much larger population of {\it small} irregular satellites than the currently known population of $R<10$-km 
irregular moons of Jupiter. This can suggest one of several things, including: (i) The population of small 
irregular moons is strongly incomplete. (ii) The original population of small moons was much larger but was 
later removed by disruptive collisions (Nesvorn\'y et al. 2003, Bottke et al. 2010). (iii) The original SFD 
did not follow a simple power law below 10 km. Instead, it was wavy.

Taken at its face value, the capture probability estimated here is somewhat inadequate to explain Jupiter's 
largest irregular moon, Himalia. This discrepancy becomes slightly smaller if the real dimensions of Himalia 
are 120 km by 150 km, as reported by Porco et al. (2003), indicating effective $R \simeq 60$-75 km (rather than 
$R \simeq 85$ km taken from the JPL Horizons site and used in Figure \ref{sfd}). This would move Himalia
closer to the model curves in Figure \ref{sfd}. Alternatively, Himalia may have been captured before the 
epoch of planetary encounters (e.g., \'Cuk \& Burns 2002) and survived on a bound orbit. We find that the 
fraction of surviving jovian moons with $0.05<a<0.1$ AU is $\sim$0.01-0.3 depending on the case considered 
(low in Case 2 and high in Cases 1 and 3). 

We may have sub-estimated $P_{\rm cap}$ by only considering encounters with $r<R_{\rm H,1}+R_{\rm H,2}$. In fact, 
as Fig. \ref{encs} shows, even encounters with $r=R_{\rm H,1}+R_{\rm H,2}\simeq0.5$ AU lead to capture of stable 
satellites. This raises the possibility that even encounters with $r>0.5$ AU can lead to satellite capture, 
which can be important, because the number of encounters increases with $r$ as $r^2$. Unfortunately, there is 
no easy way for us to include these distant encounters at this time, because this would essentially require 
to repeat the whole analysis. For reference, the original analysis required several months of computer time 
on 50 CPUs.  
\subsection{Other Planets}
NVM07 showed that the populations of irregular satellites at Saturn, Uranus and Neptune can be explained by 
capture during planetary encounters. Their model for planetary encounters, however, was based on the ONM
(Tsiganis et al. 2005), such as it is not clear whether similar results can be obtained in the new instability
cases considered here. We therefore discuss the results for these three planets below. 

\begin{figure}
\epsscale{1.0}
\plotone{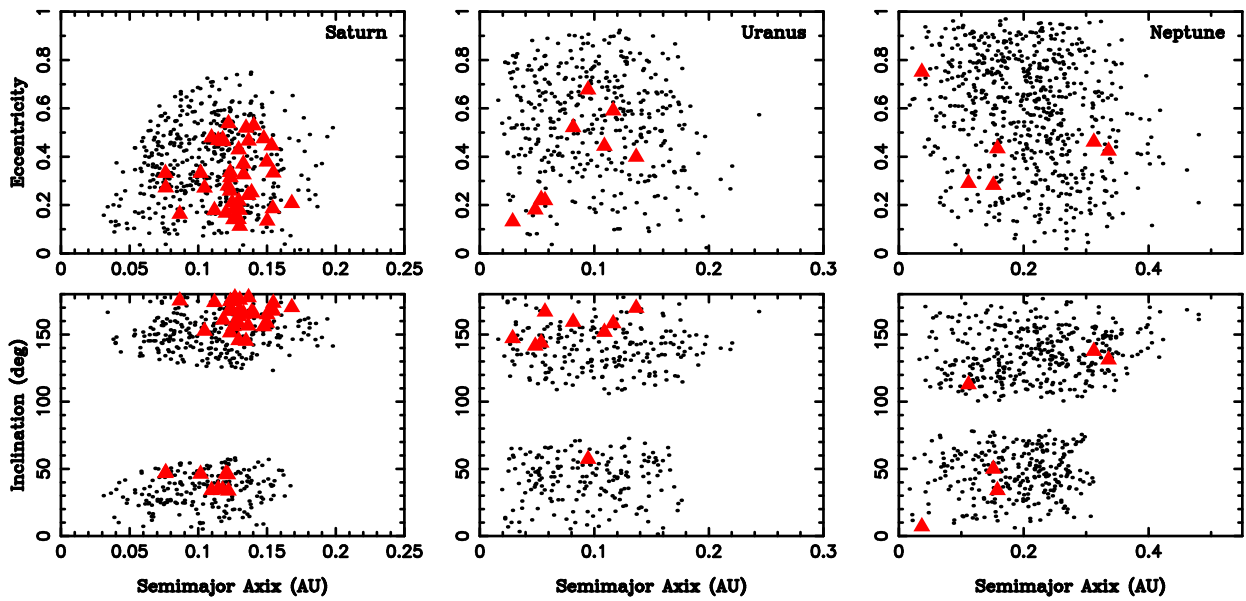}
\caption{The orbits of captured particles (dots) and known irregular satellites (red triangles).
From left to right the panels show the orbits at Saturn, Uranus and Neptune. The results for Case 
1 are shown here. The orbital distributions obtained in Cases 2 and 3 are similar.}
\label{orbits}
\end{figure}

The orbital distributions obtained in all cases considered here represent a good match to observations
(Figure \ref{orbits}). Relative to NVM07, we obtain a slightly broader semimajor axis range for satellites
captured at Neptune (compare our distributions to those shown in Figures 4 and 5 in NVM07), which is good 
to explain the distant orbits of two known retrograde moons (Psamathe and Neso). We do not consider, however,
this difference significant enough to discriminate between models. The orbital distributions of satellites
captured at Saturn and Uranus look similar to those reported in NVM07.

As for the capture probability, here we obtain $P_{\rm cap}\sim 5\times10^{-8}$ for Saturn and 
$P_{\rm cap}\simeq(1$-$3)\times10^{-8}$ for Uranus and Neptune. These values are a factor of several lower 
than those reported for Saturn, Uranus and Neptune in NVM07. This should not be a problem, however, because 
the number of satellites captured at these planets in NVM07 were factor of several larger than needed. 
These differences are related to the detailed history of encounters in the instability models. With more
encounters of Saturn, Uranus and Neptune, as in NVM07, more satellites are captured at these planets. 
Additional differences are caused by the number and radial distribution of planetesimals at the onset
of planetary encounters. 
 
A closer investigation into this issue goes beyond the scope of this paper, because it will require 
exploration of a wider range of the planetary instability models. In particular, we will need to identify 
the initial conditions such that not only all planets have encounters during the instability, but also 
that the encounters occur in exactly the right proportion.\footnote{Note that the simulations reported 
here were CPU expensive such that we were unable to more fully explore parameter space.} This effort
will help to better constrain the initial state from from which the solar system evolved. 
\section{Conclusions}
We find that planetary encounters can lead to capture of a population of irregular satellites at Jupiter
that favorably compares, both in number and orbital distribution, with the known irregular moons of Jupiter.
This resolves the problem identified in NVM07, where it was pointed out that the ONM does not provide a 
unified framework for capture of irregular satellites at all outer planets, because Jupiter does not 
typically have encounters with another planet. Using the jumping-Jupiter models from NM12, instead, appears 
to be a step in good direction, because these models allow us to extend the capture process suggested in 
NVM07 to Jupiter as well. Moreover, the capture probabilities obtained here for different planets are 
similar (to within a factor of few) explaining why the populations of irregular moons at different planets 
are roughly similar in number (Jewitt \& Haghighipour 2007). In broader context, the work presented here 
provides support for the jumping-Jupiter model (Morbidelli et al. 2009a, 2010; Brasser et al. 2009), and 
shows a good consistency of the planetary instability simulations published in NM12. 

\acknowledgments
This work was supported by NASA's Outer Planet Research program. The work of D.V. was partly supported 
by the Czech Grant Agency (grant P209-13-013085). R. D. was supported by FAPESP (grants 2012/23732 and 
2010/11109).

\clearpage
\begin{table}[t]
\begin{center}
\begin{tabular}{lrrr}
\hline 
                               & Case 1   & Case 2  & Case 3 \\
$N_{\rm enc}$                   & 59           & 280             & 49      \\
$f_{\rm 3AU}$ ($10^{-4}$)      & 4.2          & 4.0            & 3.5      \\   
$N_{\rm cap}$                   & 1458         & 3617             & 1403      \\
$P_{\rm cap}$ ($10^{-8}$)       & $1.5$        & $3.6$           & $1.3$  \\
\hline
\end{tabular}
\end{center}
\caption{The capture statistics of irregular satellites at Jupiter. The rows are: the (1) number of 
encounters ($N_{\rm enc}$), (2) mean fraction of disk particles in the 3 AU sphere around encounters 
($f_{\rm 3AU}$), (3) number of captured {\it stable} irregular satellites ($N_{\rm cap}$), and (4) probability 
of capture ($P_{\rm cap}$).}
\end{table}

\clearpage

\clearpage

\clearpage

\clearpage

\clearpage

\clearpage

\end{document}